\documentclass[a4paper]{jpconf}
\usepackage{graphicx}
\usepackage{float}
\begin{document}
\title{Measurements of $t\overline{t}$ spin correlations in CMS}

\author{Kelly Beernaert}

\address{Universiteit Gent, Ghent, Belgium}

\ead{kelly.beernaert@cern.ch}

\begin{abstract}
We present an overview of the measurements of $t\overline{t}$ spin correlations in the CMS Collaboration. We present two analyses both in the dilepton channel using proton-proton collisions at $\sqrt{s}\, =\, 7$ TeV based on an integrated luminosity of 5.0 fb$^{-1}$. The spin correlations and polarization are measured using angular asymmetries. The results are consistent with unpolarized top quarks and Standard Model spin correlation. The second analysis sets a limit on the real part of the top-quark chromo-magnetic dipole moment of $-0.043\, <\, Re({\hat{\mu}}_{t})\, <\, 0.117$ at $95\,\%$ confidence level through the measured azimuthal angle difference between the two charged leptons from $t\overline{t}$ production.
\end{abstract}

\section{Spin correlation strength}
Due to the very short top-quark lifetime, top quarks decay before their spins decorrelate. The top-quark pair spin correlation information is therefore propagated to the decay products. A measurement of the $t\overline{t}$ spin correlation strength C allows for a stringent test of the Standard Model (SM), sensitive to the spin of the top quark and the existence of new types of strong coupling.\\
The spin correlation strength C appears in the double angular distribution, 
\begin{equation}
\frac{d\sigma}{dcos\theta_{1} dcos\theta_{2}}\, =\, \frac{\sigma}{4}(1+\alpha_{1}P_{1}cos(\theta_{1}) + \alpha_{2}P_{2}cos(\theta_{2}) -Ccos(\theta_{1}) cos(\theta_{2}))  
\label{eq.doublediff}
\end{equation}
where the angle $\theta$ is the angle between the direction of the top-quark decay product in the parent rest frame and the direction of the top quark with respect to the chosen basis, P$_{1,2}$ is the top (anti-)quark polarization and $\alpha_{i}$ is the spin analyzing power.    

\section{Measurement of spin correlation and top-quark polarization}

The presented measurement of spin correlation and top-quark polarization was performed in the dilepton channel at $\sqrt{s}\, =\, 7$ TeV using 5.0 fb$^{-1}$ of integrated luminosity \cite{PhysRevLett.112.182001}, by means of three asymmetries shown below. The first asymmetry, in eq.\,\ref{eq.pol}, accesses the top-quark polarization P (assuming P$_{1}$ = P$_{2}$ = P) with $P\,=\,2A_{P}$ in the helicity basis 
\begin{equation}
A_{p}\, = \, \frac{N[cos(\theta^{*}_{l}) > 0] - N[cos(\theta^{*}_{l}) < 0]}{N[cos(\theta^{*}_{l}) > 0] + N[cos(\theta^{*}_{l}) < 0]}
\label{eq.pol}
\end{equation}
counting both the positive and negative leptons, where $\theta^{*}_{l}$ is the angle of a charged lepton in the rest frame of its parent top quark or antiquark, measured in the helicity frame. The variable $A_{\Delta \phi}$ provides discrimination between correlated and uncorrelated $t\overline{t}$ spins with $\Delta \phi$ the difference in azimuthal angle between the charged leptons

\begin{equation}
A_{\Delta \phi}\, =\, \frac{N(\Delta \phi_{l^{+}l^{-}} > \pi/2) - N(\Delta \phi_{l^{+}l^{-}} < \pi/2)}{N(\Delta \phi_{l^{+}l^{-}} > \pi/2) + N(\Delta \phi_{l^{+}l^{-}} < \pi/2)}
\label{eq.azim}
\end{equation}
.
The third asymmetry value, given in eq.\,\ref{eq.spin}, is a direct measure of the spin correlation strength with $C_{hel}\,=\, -4A_{c_{1}c_{2}}$ using the helicity angle of the leptons with $c_{1}\,=\,cos(\theta^{*}_{l^{+}})$ and $c_{2}\,=\,cos(\theta^{*}_{l^{-}})$

\begin{equation}
A_{c_{1}c_{2}}\, =\, \frac{N(c_{1}c_{2} > 0) - N(c_{1}c_{2} < 0)}{N(c_{1}c_{2} > 0) + N(c_{1}c_{2} < 0)}
\label{eq.spin}
\end{equation}
.

\begin{figure}[H]
\begin{center}
\begin{tabular}{cc}
\includegraphics[width=0.44\textwidth]{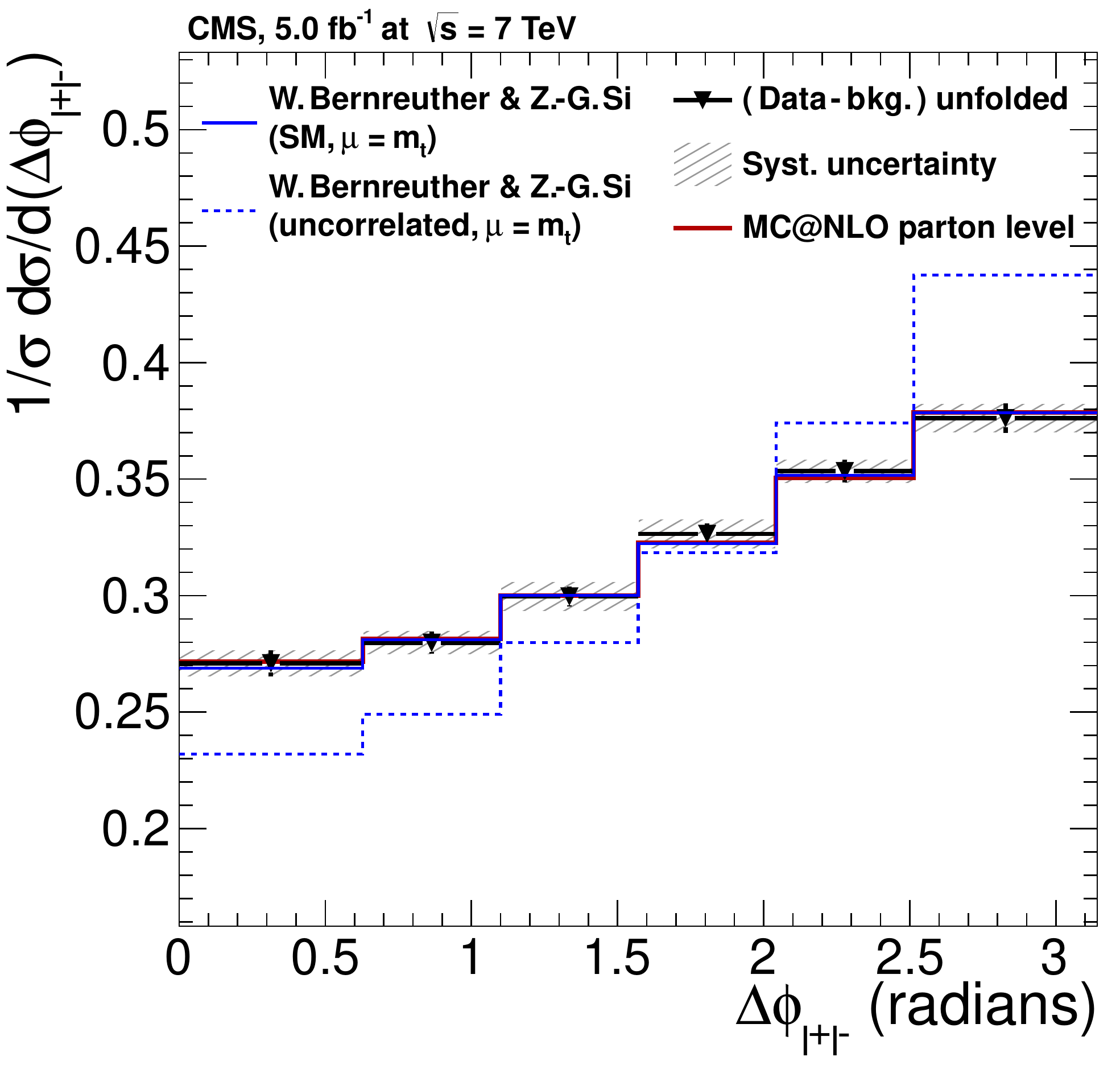} &
\includegraphics[width=0.44\textwidth]{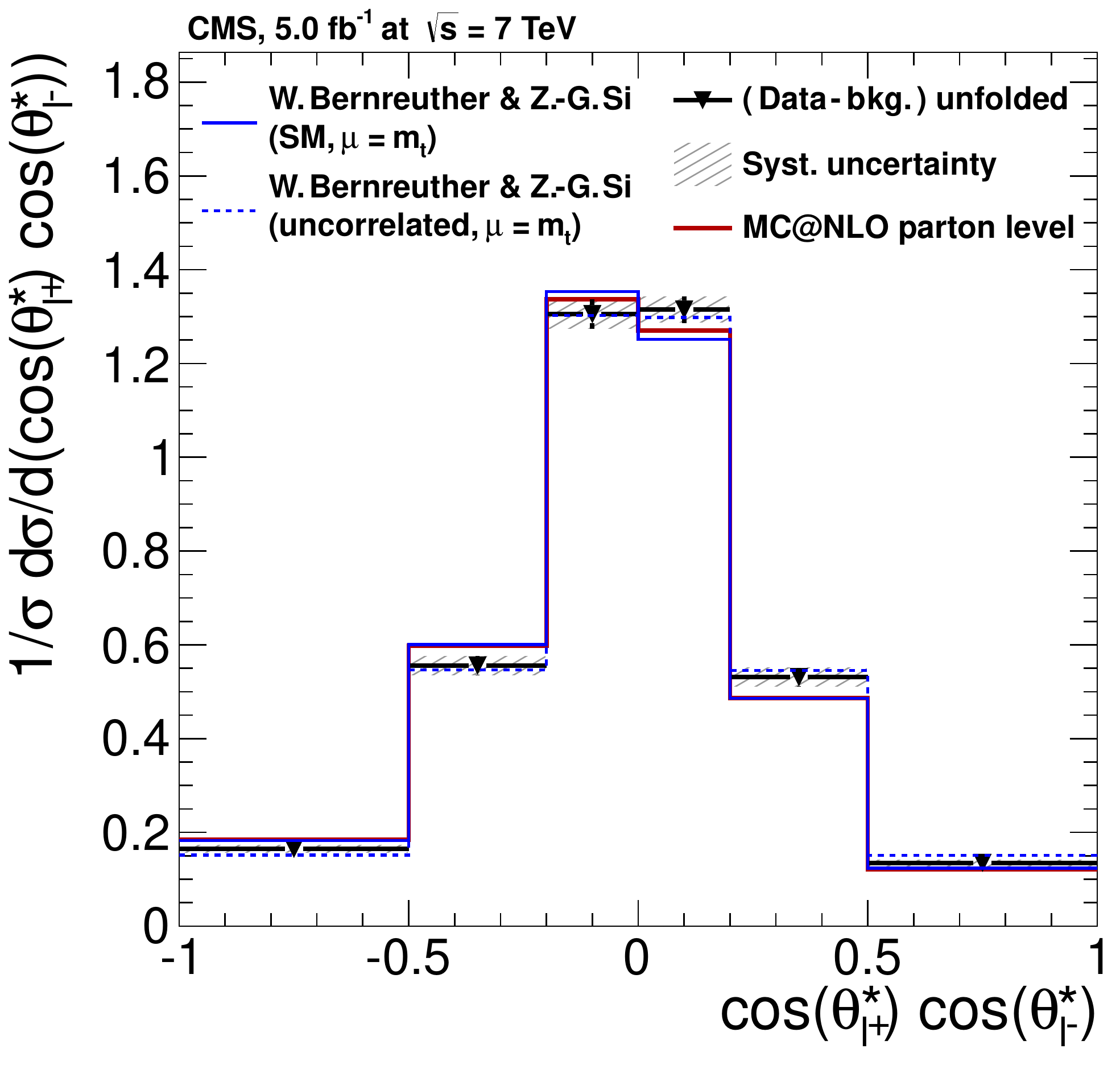} \\
\includegraphics[width=0.44\textwidth]{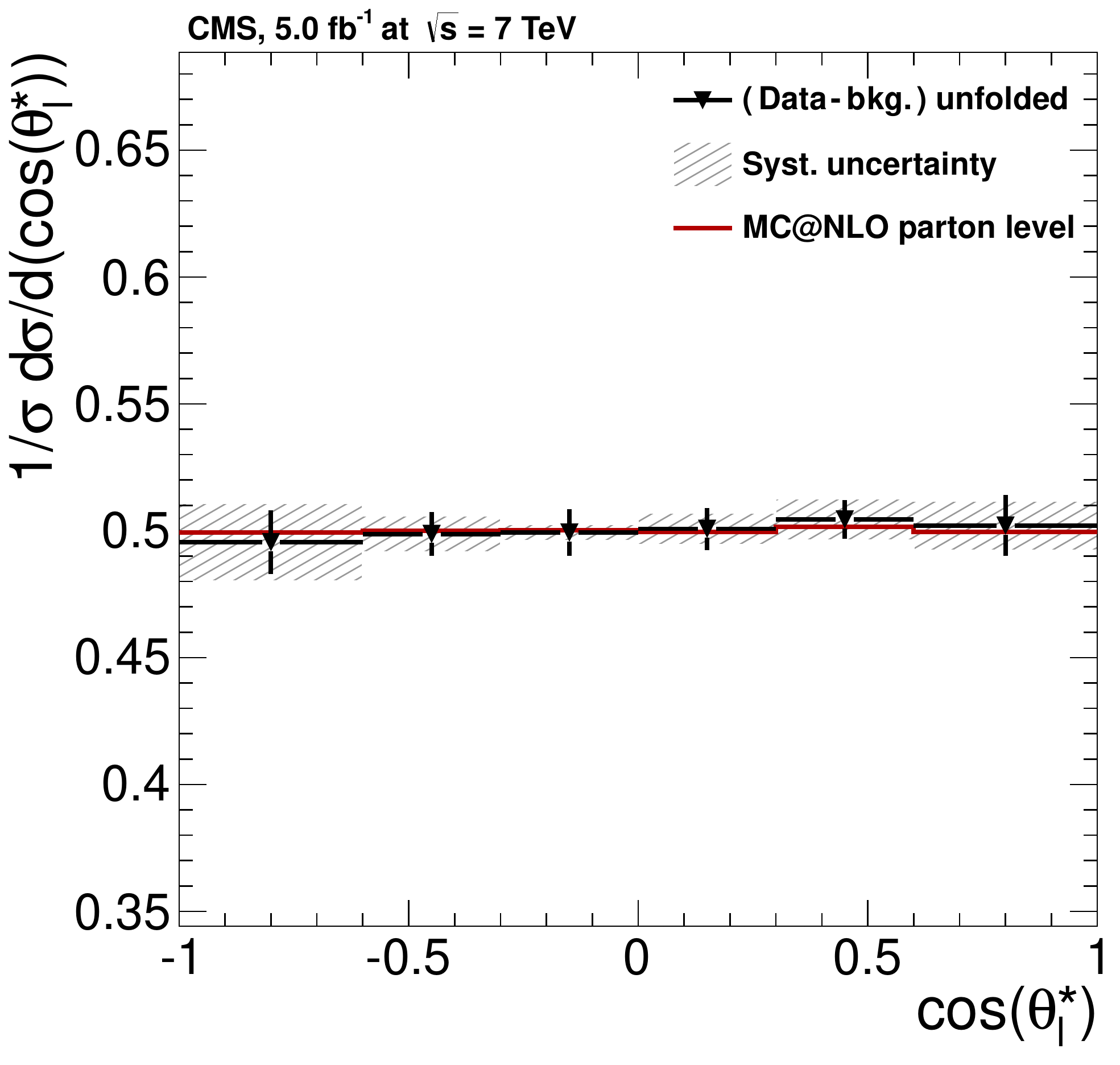} & \\
\end{tabular}
\end{center}
\caption{Background-subtracted and unfolded distributions for $\Delta \phi_{l^{+}l^{-}}$ , $cos(\theta^{*}_{l^{+}} )cos(\theta^{*}_{l^{-}} )$ and $cos(\theta^{*}_{l})$. The error bars are statistical only, the hatched area is the systematic uncertainty band. Due to unfolding the bin contents are correlated.}
\label{fig:Plots}
\end{figure}

\begin{table*}[!htpb]
\centering
\caption{\label{tab:ResultsUnfolded}
Parton-level asymmetries. The uncertainties in the unfolded results are statistical, systematic, and the additional uncertainty from the top-quark $p_{t}$ reweighting.
The uncertainties in the simulated results are statistical only, while the uncertainties in the NLO calculations for correlated and uncorrelated $t\bar{t}$ spins come from scale variations up and down by a factor of two.
The prediction for $A_{c_{1}c_{2}}$ is exactly zero in the absence of spin correlations by construction.
}
\resizebox{\linewidth}{!}
{
\begin{tabular}{lrrrr}
Asymmetry  &   \multicolumn{1}{c}{Data (unfolded)} &  \multicolumn{1}{c}{MC@NLO} & \multicolumn{1}{c}{NLO (SM, correlated)} & \multicolumn{1}{c}{NLO (uncorrelated)} \\
\hline & & & & \\ [-2.0ex]
$A_{\Delta\phi}$ &   $0.113 \pm 0.010  \pm 0.006 \pm 0.012$  &  $ 0.110 \pm 0.001$  &  $0.115^{+0.014}_{-0.016}$  &  $0.210^{+0.013}_{-0.008}$    \\ [0.5ex]
$A_{c_{1}c_{2}}$ & $-0.021 \pm 0.023 \pm 0.025 \pm 0.010$  &  $ -0.078 \pm 0.001$  &  $-0.078 \pm 0.006$ &  $0$ \\ [0.5ex]
$A_{P}$ & $0.005 \pm 0.013 \pm 0.014 \pm 0.008$  & $0.000 \pm 0.001$ & ... & ... \\
\end{tabular}
}
\end{table*}

The data has been background-subtracted and the distributions are unfolded to parton-level as shown in fig.\,\ref{fig:Plots}. The results are quoted in table \ref{tab:ResultsUnfolded}. The results are compared to the values obtained in the MC@NLO simulation and to the values predicted by theory calculations to NLO \cite{Nucl.PhysB837}. The major systematic uncertainties are due to the JES uncertainty and the uncertainty due to the unfolding procedure. The uncertainty due to top-quark $p_{t}$ reweighting has been quoted separately since the origin of the effect is not yet fully understood.   

\section{Limits on top-quark chromo-magnetic moments}

Several new physics models predict an effect on the top-quark pair spin correlation strength. Anomalous couplings in the $t\overline{t}g$ vertex can be accessed via a model-independent search using an effective Lagrangian parametrised using chromo-magnetic and chromo-electric Dipole Moments (CMDM and CEDM) \cite{PhysLettB}. In an extension of the previous analysis, the differential cross section distribution is used to put limits on this effective model \cite{CMS-PAS-TOP-14-005}. The model is described by an effective Lagrangian of the following form 

\begin{equation}
L_{eff}\,=\, L_{SM} - \frac{g_{s}{\hat{\mu}}_{t}}{2m_{t}}\overline{t}\sigma^{\mu\nu}T^{a}tG^{a}_{\mu\nu} - \frac{g_{s}{\hat{d}}_{t}}{2m_{t}}\overline{t}i\sigma^{\mu\nu}\gamma_{5}T^{a}tG^{a}_{\mu\nu}
\label{eq.BSMmodel}
\end{equation}

with ${\hat{\mu}}_{t}$ and ${\hat{d}}_{t}$ the CMDM (CP-conserving) and CEDM (CP-violating) dimensionless, complex parameters. These parameters are assumed to be constant. $g_{s}$ is the strong coupling constant, $G^{a}_{\mu\nu}$ the gluon field strength, $T^{a}$ the QCD fundamental generators and $\sigma_{\mu\nu}=i(\gamma_{\mu}\gamma_{\nu}-\gamma_{\nu}\gamma_{\mu})/2$. The differential cross section $( 1/\sigma )( d\sigma/d | \Delta\phi_{ll} |)$ is sensitive to the real part of the CMDM parameter $Re({\hat{\mu}}_{t})$. From $t\overline{t}$ cross section measurements, there are already constraints on the parameters \cite{PhysLettB}. As such we can expand the differential cross section into the SM and New Physics (NP) contribution (see eq.\,\ref{eq.fitmodel}) since $Re({\hat{\mu}}_{t}) << 1$  is valid.

\begin{equation}
\frac{1}{\sigma}\frac{d\sigma}{d|\Delta\phi_{ll}|}\, =\, (\frac{1}{\sigma}\frac{d\sigma}{d|\Delta\phi_{ll}|})_{SM} + Re({\hat{\mu}}_{t})(\frac{1}{\sigma}\frac{d\sigma}{d|\Delta\phi_{ll}|})_{NP} 
\label{eq.fitmodel}
\end{equation}

\begin{figure}[h!]
\begin{center}
\begin{tabular}{cc}
\includegraphics[width=0.45\textwidth]{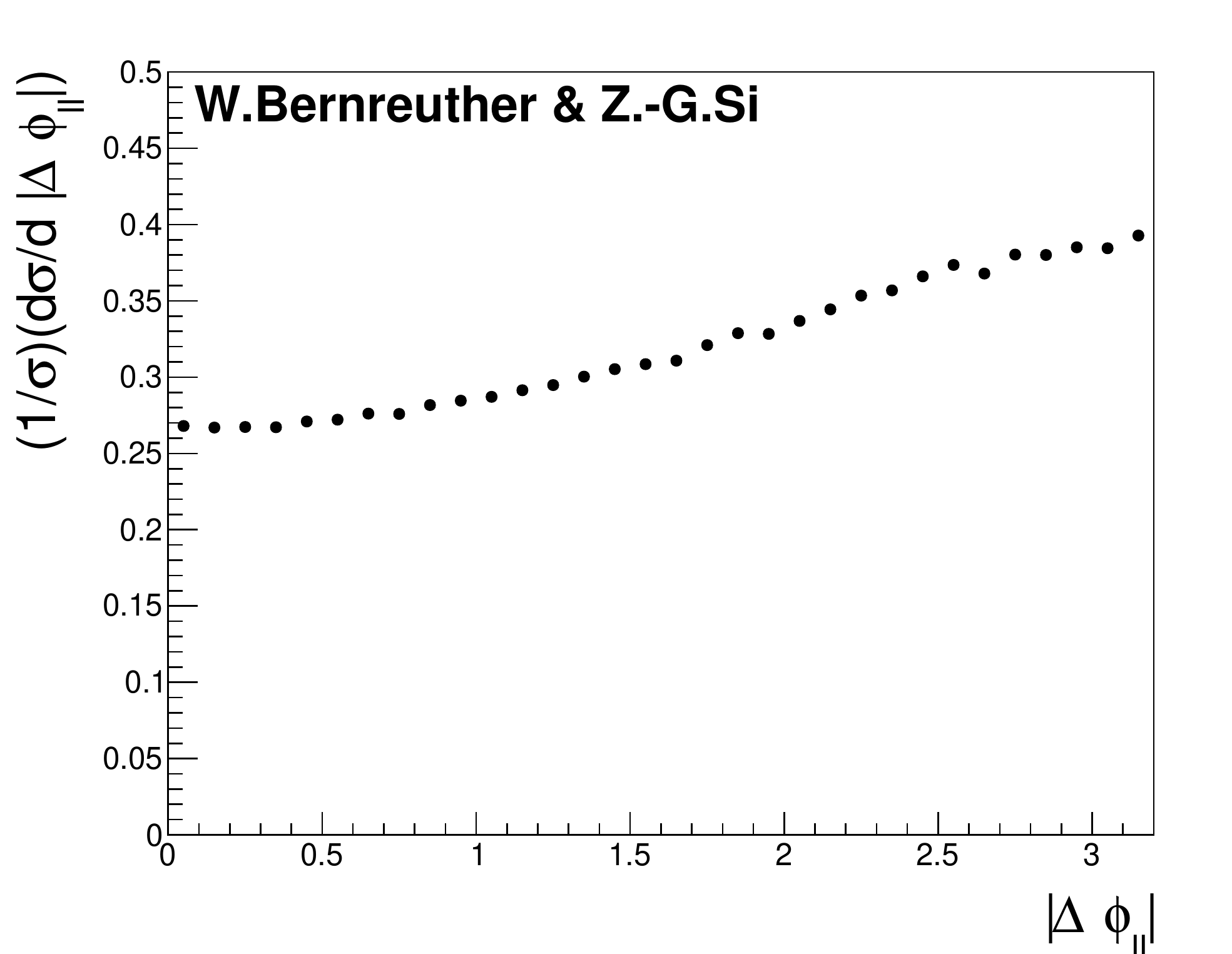} & 
\includegraphics[width=0.45\textwidth]{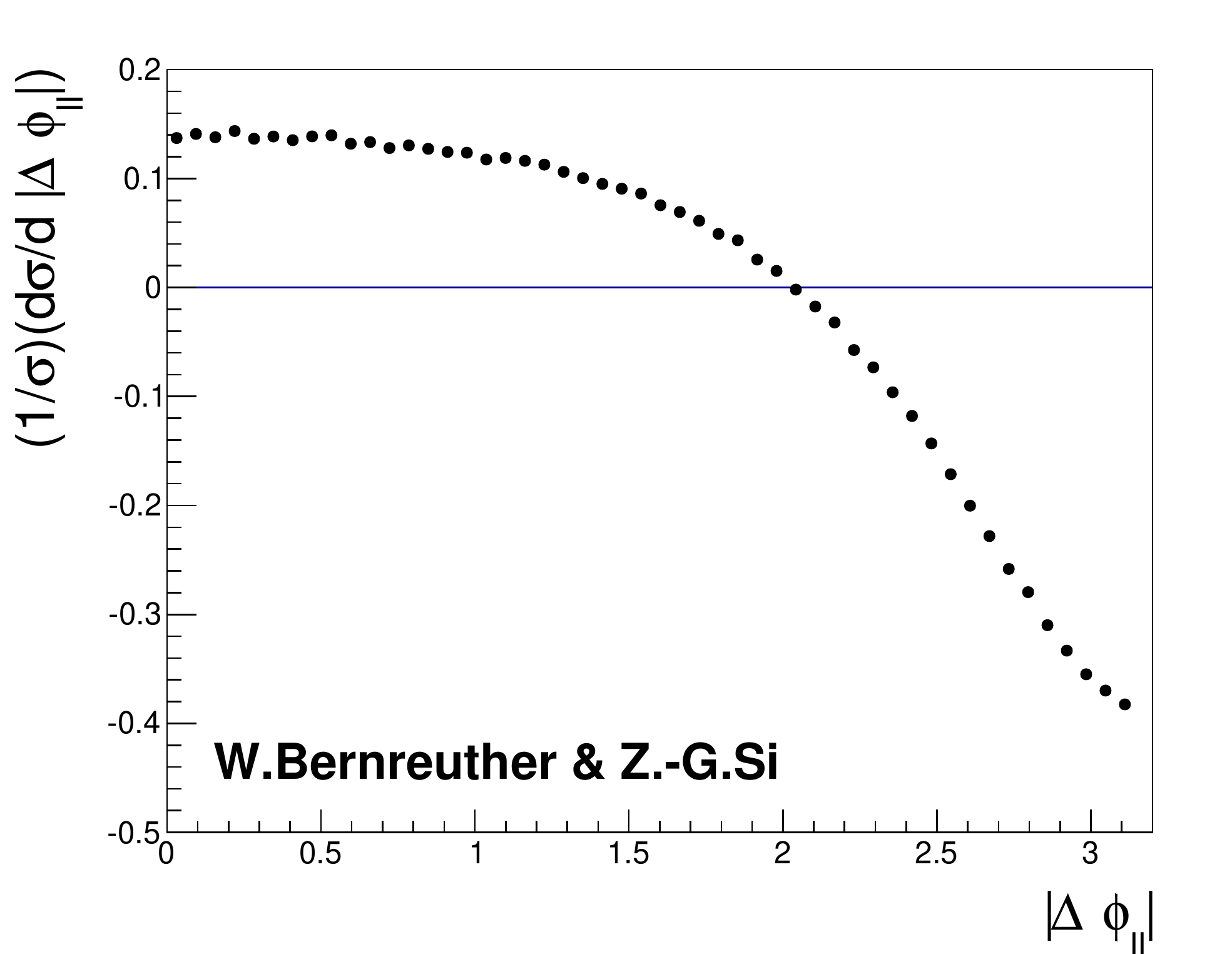}\\
\end{tabular}
\end{center}
\caption{The NLO QCD and weak (NLOW) contributions are shown from the SM (left) and the NP contribution with $Re({\hat{\mu}}_{t})\, =\, 1$ (right) at the LHC for $\sqrt{s}\, =\, 7$ TeV \cite{PhysLettB}.}
\label{fig:Bern}
\end{figure}

In fig.\,\ref{fig:Bern}, the contributions of the SM and the NP is shown, as calculated by Bernreuther and Si in \cite{PhysLettB}. The dominant theoretical uncertainty on the SM NLOW predictions is due to the choice of renormalisation and factorisation scale, while the impact of this on the NP contribution is negligible. Eq.\,\ref{eq.fitmodel} can be transformed into a fit function where $Re({\hat{\mu}}_{t})$ is the only free parameter. 

\begin{figure}[h!]
\begin{center}
\begin{tabular}{c}
\includegraphics[width=0.5\textwidth]{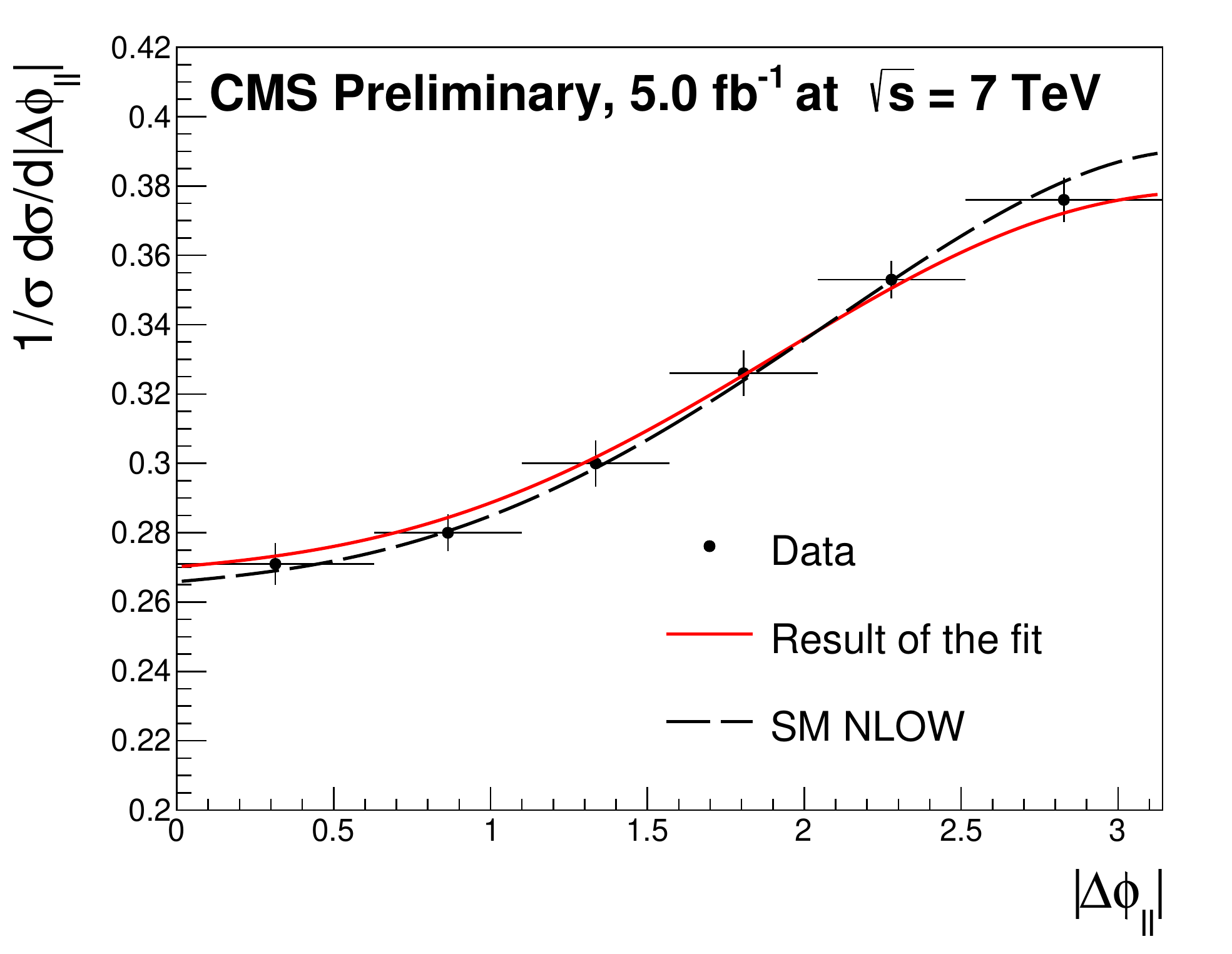}\\
\end{tabular}
\end{center}
\caption{Result of the fit of the measured $( 1/\sigma )( d\sigma/d | \Delta\phi_{ll} |)$ differential cross section.}
\label{fig:results}
\end{figure}

The shapes in fig.\,\ref{fig:Bern} are fitted by polynomials and are kept fixed in the fit to the data. The result of the fit and the SM NLOW contribution are shown in fig.\,\ref{fig:results}.

The fitted value is $Re({\hat{\mu}}_{t})\,=\,0.037\,\pm\,0.041$, after calibration and taking into account the full covariance matrix and the systematic uncertainty due to scale variations. Assuming a Gaussian behaviour of $Re({\hat{\mu}}_{t})$, the exclusion limits at $95\,\%$ C.L. are $-0.043\,<\,Re({\hat{\mu}}_{t})\,<\,0.117$.  

\section{Conclusions}
A measurement of tt spin correlations in pp collisions at $\sqrt{s}$=7 TeV has been presented. 
The data were recorded by the CMS experiment using an integrated luminosity of 5.0 fb$^{-1}$ in the dilepton channel.
The measurement of three asymmetries shows no deviations to the SM predictions for 
spin correlations in $t\overline{t}$ pair production and decay and top polarization. 
The differential cross section 
$(1/\sigma_{t\overline{t}} )(d\sigma_{t\overline{t}} /d|\Delta\phi_{ll} |)$ is used to set limits on possible contributions 
due to the top-quark 
CMDM, with this particular cross section sensitive to the real part $Re(\hat{\mu})$. This new 
physics contribution was measured to be compatible with zero and limits have been set.
\section*{References}

\bibliographystyle{plain}
\bibliography{PosterProceedings}
\end{document}